\documentclass[11pt,twoside]{article}

\usepackage{asp2004}
\usepackage{epsf}
\usepackage{lscape}
\usepackage{graphicx}

\begin{document}
\title{Hot subluminous O stars from the SDSS}   
\author{Heiko A. Hirsch,$^1$ Uli Heber,$^1$ and Simon J. O'Toole$^2$}
\affil{$^1$Dr. Remeis-Sternwarte Bamberg, Astronomisches Institut der Universit\"at Erlangen, Sternwartstrasse 7, 96049 Bamberg, Germany\\
$^2$Anglo-Australian Observatory, PO Box 296 Epping NSW 1710, Australia}    

\begin{abstract} 
We carried out a quantitative spectral analysis of 73 hot subluminous O-stars selected from the SDSS spectral database.
While the helium deficient sdOs are scattered over a wide range of effective temperature and gravity, the helium enriched sdO stars are concentrated in a small intervall of 40\,kK to 50\,kK and $\log g = 5.5 \ldots 6.0$. 
Comparing the distribution in the $T_{\rm eff}$-$\log g$-diagram with evolutionary tracks, we find the helium deficient sdOs to be the progeny of the sdB stars.
The results for the helium enriched ones are less conclusive.
Both the merger of two white dwarfs and the delayed helium core flash scenarios are viable options to be explored further.
\end{abstract}

\section*{Introduction}
Hot subluminous stars can roughly be divided into two classes: The hotter and in general helium rich subluminous O stars (sdO) and the helium deficient subluminous B stars (sdB).
Following recent work of \citet*{stro06}, we divide the sdOs in helium enriched and helium deficient sdOs, depending on whether they show super- or subsolar helium abundances.

The sdB stars have been identified as extended horizontal branch stars (EHB).
They form a quite homogenous group of helium core burning stars with a very thin ($< 0.02\,M_\odot)$ hydrogen shell \citep*{hebe86}.
The unsolved question is, how the high mass loss necessary for a star to reach the EHB is accomplished.

SdO stars on the other hand are less well understood.
The sdOs are often thought of as the progeny of the sdB stars.
A recent study by \citet{stro06} finds effective temperature ranges from 40\,000\,K up to $\approx$80\,000\,K, the gravity from $\log g = 4.8 \ldots 6.5$ (cgs) and helium abundance from $\log (N_{He} / N_H) = -3 \ldots +3$, helium enriched stars are frequent.
This poses a difficult problem for canonical evolution theory.
It is quite hard to explain, how helium deficient sdB stars can evolve into highly helium enriched sdOs.

Our aim is to better understand the origin of sdOs and to search for possible links between them and the sdBs.

\section*{Spectral analysis}
The huge spectral database of the SDSS 
provides spectra with a moderate resolution of $R \approx 1800$ from 3800 to 9200\,{\AA}.
To study a large sample of sdO stars, we started an extensive search by selecting all stars within the colour range $(u-g)<0.4$ and $(g-r)<0.1$ in the Data Release 4.
Eventually we found spectra of 73 sdOs useful for a quantitative spectral analysis.

We performed a fit of synthetic spectra from state-of-the-art NLTE model atmospheres to the data using a $\chi^2$-routine \citep*{napi99}.
The model atmospheres contain hydrogen and helium only and take into account partial line blanketing.
68\% of all stars have errors better than $\Delta T_{\rm eff} = 1\,500\,K$, $\Delta\log g = 0.18$ and $\Delta\log (N_{He} / N_H) = 0.14$.
We combine our results with those for the 56 sdOs by \citet{stro06}, in order to improve statistics.
The stars are widely scattered in $T_{\rm eff}$ and $\log g$ and no sdO is located on the EHB, see Fig.~\ref{image1}.
The helium enriched sdOs are concentrated in a narrow interval of $T_{\rm eff} = 40\,000 \ldots 50\,000\,K$, $\log g = 5.2 \ldots 6.3$, while the helium deficient ones avoid this region.
It is worthwhile to note that a number of sdOs lie below the helium zero age main sequence (He-ZAMS).

\section*{Stellar Evolution}

\begin{figure}
\includegraphics[width=0.49\textwidth]{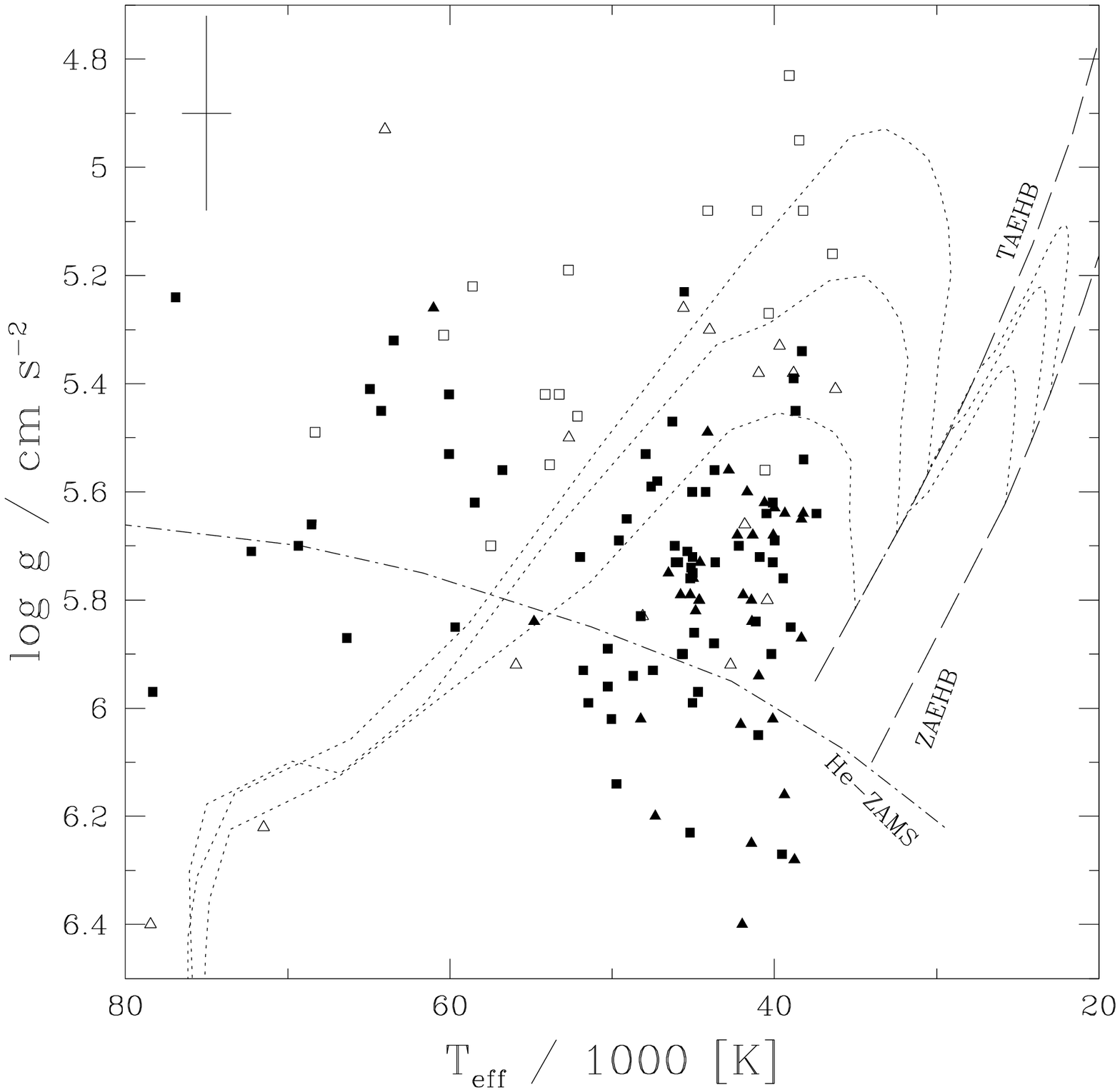}
\includegraphics[width=0.49\textwidth]{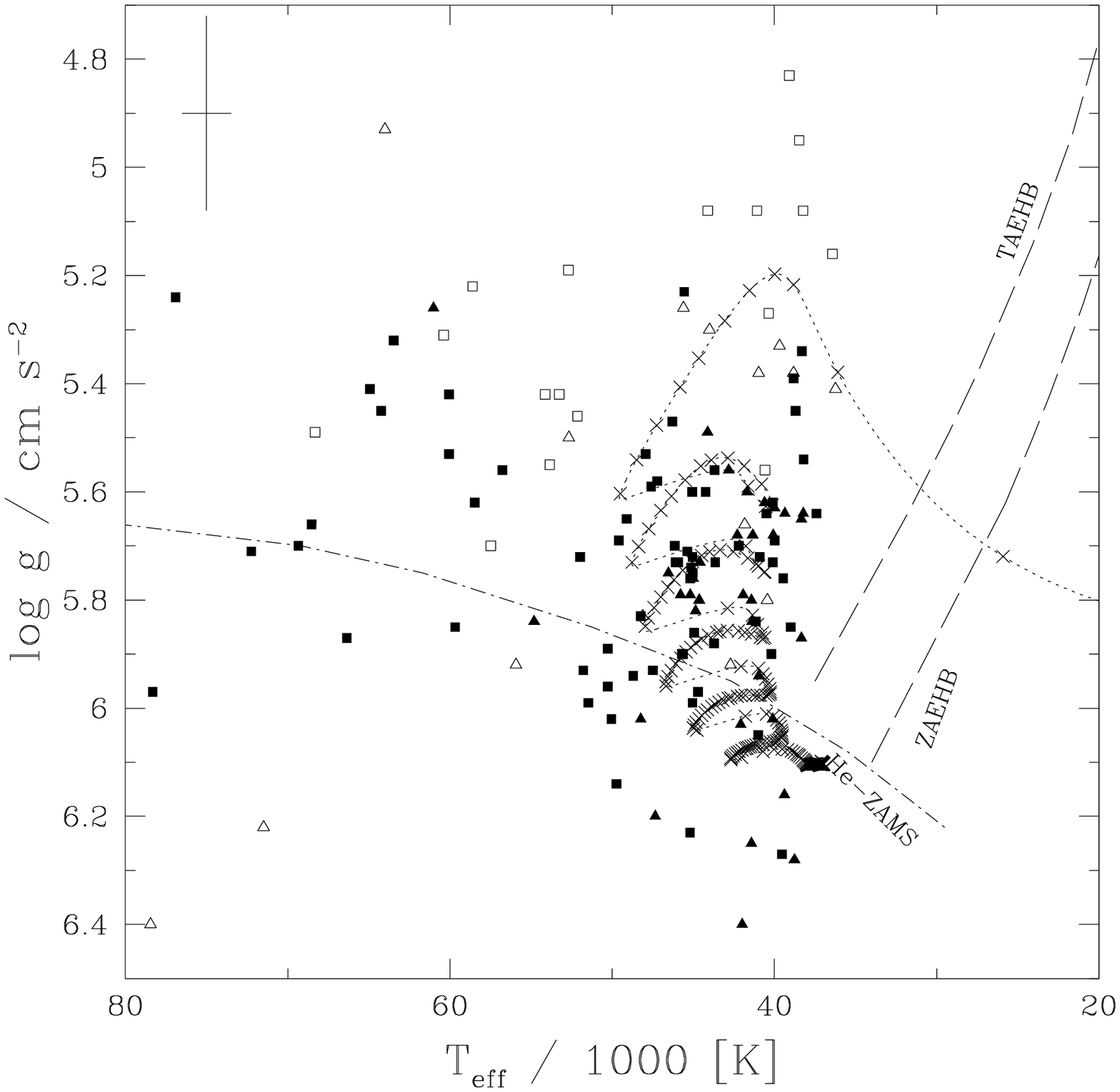}
\caption {$T_{\rm eff}$-$\log g$-diagram of sdO stars from SDSS (squares) and from SPY \citep{stro06} (triangles).
The filled symbols represent helium enriched stars, the open symbols helium deficient ones.
The zero age EHB (ZAEHB) and the terminal age EHB (TAEHB) as well as the helium zero age main sequence (He-ZAMS) are plotted.
68\% of the stars have errors smaller than given in the upper left corner of each panel.
\textit{Left panel:} Paths shown are post-EHB evolution tracks by \citet{dorm93} for 0.471, 0.473 and 0.475\,$M_\odot$ core masses (from bottom to top).
\textit{Right panel:} The evolutionary path of a late hot flasher is plotted.
Crosses mark equidistant time steps of 10\,000 years, revealing a significant discrepancy of evolutionary time scales and the observed distribution.}
\label{image1}
\end{figure}

Canonical evolution theory sees the sdOs as progeny of the sdB stars.
\citet*{dorm93} have calculated appropriate tracks, evolving a star from the zero age EHB (ZAEHB) to the terminal age EHB (TAEHB) and then through the sdO regime to higher temperatures and higher surface gravities, see Fig.~\ref{image1} (left panel).
They fail to explain the clustering of helium enriched sdOs, but reproduce the population of helium deficient sdOs well.
Hence \textbf{the helium deficient stars form the progeny of the sdB stars} while the helium enriched ones do not.

The problems we face in linking the helium enriched sdOs to the sdBs is the cause of the helium enrichment and the clumping in the ($T_{\rm eff}$, $\log g$) domain.
Therefore we investigate the scenario proposed by \citet*{swei97} for single star evolution:

\textbf{Late hot flashers} are stars that undergo the helium core flash not until they are already descending the white dwarf cooling curve.
The delayed flash induces mixing which will transport hydrogen into the core, resulting in a helium burning star with $T_{\rm eff} \approx 40\,000\,K$ on or near the He-ZAMS, enriched with carbon or nitrogen.
An example of this evolution is shown in Fig.~\ref{image1} (right panel) for a star with a resulting composition of $X = 0.154, Y=0.814$.
It provides a promising explanation for the clustering of helium enriched sdO stars, but our observed population does not match the predicted evolutionary timescales.
We would expect many more stars near the He-ZAMS than observed.

An alternative scenario to explain the helium enriched sdOs is close binary evolution.
Three channels leading to a helium burning star with a tiny hydrogen envelope have been discussed by \citet{han02}:
Such stars can be formed by stable Roche Lobe overflow, common envelope ejection and the merging of two helium white dwarfs.

The \textbf{merger scenario} is a promising option to explain sdO stars.
Short period binaries of two helium white dwarfs will loose energy through radiation of gravitational waves.
Eventually, they will merge and the resulting star will ignite helium.
\citet{han02} argue, that this merging process will mix the hydrogen shells into the helium burning regions where it will rapidly be consumed and bring up processed material to the surface.

\begin{figure}
\begin{center}
\includegraphics[width=0.49\textwidth]{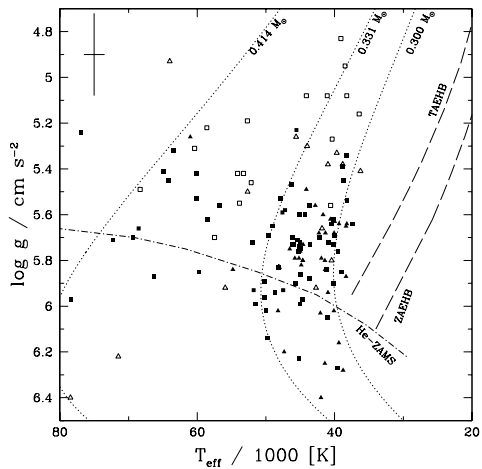}
\caption{$T_{\rm eff}$-$\log g$-diagram. Symbols as in Fig.~\ref{image1}. Plotted are post-RGB tracks by \citet*{drie98}.}
\label{image2}
\end{center}
\end{figure}

\textbf{Post-RGB evolution} is another possible origin for hot subdwarfs.
A giant star can leave the RGB before the onset of helium burning, if mass transfer in the RGB phase took place, resulting in a helium star with a thin hydrogen envelope left \citep*{drie98}.
Hence such a star would not burn helium but evolve through the sdO domain towards a helium core white dwarf.
Such tracks are plotted in Fig.~\ref{image2}.
They cover the helium enriched sdO regime quite well, including the otherwise hard to explain stars below the He-ZAMS.

However, \citet{drie98} find a He/H mass ratio of $Y = 0.3$ in the relevant regime, an order of magnitude lower than most helium enriched sdOs posses.
In addition, while these stars should have a close companion, \citet{napi04} find only one out of 23 helium enriched sdO stars to be radial velocity variable.
Therefore, this scenario is unlikely to be relevant for sdO formation.

\section*{Summary and conclusion}
We did a search for spectra of subluminous O stars in the SDSS database and performed a spectral analysis for 73 stars.
The analysis revealed a clustering of helium enriched sdOs in a small intervall at about 45\,000\,K, a region almost void of the helium deficient stars.
A comparison of the sdOs in the $T_{\rm eff}$-$\log g$-diagram with different evolutionary tracks showed that the helium deficient sdO stars probably form the progeny of the sdB stars.
The stars with helium enriched atmospheres still pose a problem: both the late hot flasher and the white dwarf merger scenario can explain the observations qualitatively, but fail to do so in the detail.
For further analysis we have started a project to measure carbon and nitrogen abundances which will yield tight constraints on the formation process.

\acknowledgements 
We like to thank the \emph{Royal Astronomy Society} for their generous grant and the \emph{Deutsche Forschungs Gemeinschaft} for their support.

\end{document}